# Quantum in the Cloud: Application Potentials and Research Opportunities

Frank Leymann[a], Johanna Barzen[b], Michael Falkenthal[c],
Daniel Vietz[d], Benjamin Weder[e], Karoline Wild[f]

*Institute of Architecture of Application Systems, University of Stuttgart, Universitätsstr. 38, Stuttgart, Germany*
*{firstname.lastname }@iaas.uni-stuttgart.de*

Abstract: Quantum computers are becoming real, and they have the inherent potential to significantly impact many application domains. We sketch the basics about programming quantum computers, showing that quantum programs are typically hybrid consisting of a mixture of classical parts and quantum parts. With the advent of quantum computers in the cloud, the cloud is a fine environment for performing quantum programs. The tool chain available for creating and running such programs is sketched. As an exemplary problem we discuss efforts to implement quantum programs that are hardware independent. A use case from machine learning is outlined. Finally, a collaborative platform for solving problems with quantum computers that is currently under construction is presented.

## 1 INTRODUCTION

Quantum computing advanced up to a state that urges attention to the software community: problems that are hard to solve based on classical (hardware and software) technology become tractable in the next couple of years (National Academies, 2019). Quantum computers are offered for commercial use (e.g. IBM Q System One), and access to quantum computers are offered by various vendors like Amazon, IBM, Microsoft, or Rigetti via the cloud.

However, todays quantum computers are error-prone. For example, the states they store are volatile and decay fast (decoherence), the operations they perform are not exact (gate fidelity) etc. Concequently, they are "noisy". And their size (measured in Qubits – see section 2.1) is of "intermediate scale". Together, todays quantum computers are Noisy Intermediate Scale Quantum (NISQ) computers (Preskill, 2019). In order to perform a quantum algorithm reliably on a NISQ machine, it must be limited in size.

Because of this, the overall algorithms are often hybrid. They perform parts on a quantum computer, other parts on a classical computer. Each part performed on a quantum computer is fast enough to produce reliable results. The parts executed on a classical computer analyze the results, compute new parameters for the quantum parts, and pass them on to a quantum part. Typically, this is an iteration consting of classical pre-processing, quantum processing, and classical post-processing.

This iteration between classical parts and quantum parts reveals why the cloud is a solid basis for executing quantum applications: it offers classical environments as well as quantum computers (see before).

What are viable applications on NISQ computers? For example, simulation of molecules in drug discovery or material science is very promising (Grimsley et al., 2019), many areas of machine learning will realize significant improvements (Dunjko et al., 2016), as well as solving optimization problems (Guerreschi et al., 2017).

[a] https://orcid.org/0000-0002-9123-259X
[b] https://orcid.org/0000-0001-8397-7973
[c] https://orcid.org/0000-0001-7802-1395
[d] https://orcid.org/0000-0003-1366-5805
[e] https://orcid.org/0000-0002-6761-6243
[f] https://orcid.org/0000-0001-7803-6386

## 1.1 Paper Overview

Section 2 sketches the programming model of quantum computers. Quantum computing in the cloud is introduced in section 3. How to remove hardware dependencies is addressed in section 4. Section 5 outlines a use case of quantum machine learning. A collaboration platform for developing and exploiting quantum applications is subject of section 6. Section 7 concludes the paper.

## 2 PROGRAMMING MODEL

Next, we introduce the basics of the quantum programming model – see (Nielsen et al., 2016).

### 2.1 Quantum Registers

The most fundamental notion of quantum computing is the quantum bit or qubit for short. While a classical bit can have either the value 0 or 1 at a given time, the value of a qubit $|x\rangle$ is any combination of these two values: $|x\rangle=\alpha\cdot|0\rangle+\beta\cdot|1\rangle$ (to distinguish bits from qubits we write $|x\rangle$ instead of x for the latter). This so-called superposition is one source of the power of quantum computing.

The actual value of a qubit is determined by a so-called measurement. $\alpha^2$ and $\beta^2$ are the probabilities that – once the qubit is measured – the classical value "0" or "1", respectively, results. Because either "0" or "1" will definitively result, the probabilities sum up to 1: $\alpha^2+\beta^2=1$.

Just like bits are combined into registers in a classical computer, qubits are combined into quantum registers. A quantum register $|r\rangle$ consisting of n qubits has a value that is a superposition of the $2^n$ values $|0...0\rangle$, $|0...01\rangle$, up to $|1...1\rangle$. A manipulation of the quantum register thus modifies these $2^n$ values at the same time: this quantum parallelism is another source of the power of quantum computing.

### 2.2 Quantum Operations

Figure 1 depicts two qubits $\alpha|0\rangle+\beta|1\rangle$ and $\gamma|0\rangle + \delta|1\rangle$: because $\alpha^2+\beta^2 = \gamma^2+\delta^2 = 1$, each qubit can be represented as a point on the unit circle, i.e. as a vector of length 1. Manipulating a qubit results in another qubit, i.e. a manipulation U of qubits preserves the lengths of qubits as vectors. Such manipulations are called unitary transformations. A quantum algorithm combines such unitary transformations to manipulate qubits (or quantum registers in general). Since the combination of unitary transformations is again a unitary transformation, a quantum algorithm is represented by a unitary transformation too.

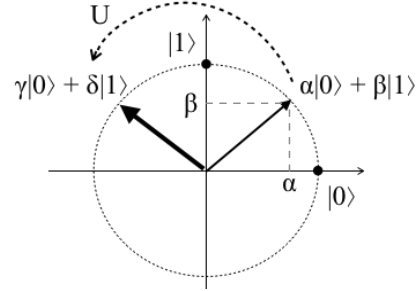

Figure 1: Depicting a qubit and its manipulation.

This geometric interpretation of qubits is extended to quantum registers: a quantum register with n qubits can be perceived as a unit vector in a $2^n$-dimensional vector space. A quantum algorithm is then a unitary transformation of this vector space.

A quantum algorithm U takes a quantum register $|r\rangle$ as input and produces a quantum register $|s\rangle=U(|r\rangle)$ as output, with

$$|s\rangle = \sum_{i=1}^{2^n} \alpha_i |x_1^i \cdots x_n^i\rangle, x_j^i \in \{0,1\} \quad (1)$$

The actual result of the algorithm U is determined by measuring $|s\rangle$. Thus, the result is $(x_1^i \cdots x_n^i) \in \{0,1\}^n$ with probability $\alpha_i^2$. Obviously, different executions of U followed by a measurement to determine U's result will produce different bit-strings according to their probability: A single execution of a quantum algorithm is like a random experiment. Because of this, a quantum algorithm is typically performed many times to produce a probability distribution of results (see Figure 2 for an example) – and the most probable result is taken as "the" result of the quantum algorithm.

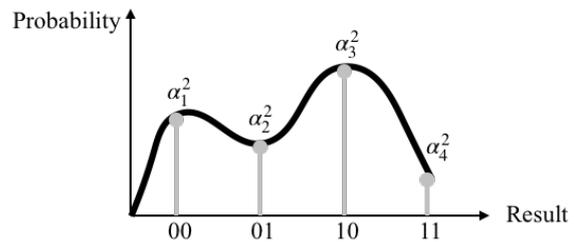

Figure 2: Depicting a qubit and its manipulation.

## 2.3 Quantum Algorithms

As shown in Figure 3, the core of a quantum algorithm is a unitary transformation – which represents the proper logic of the algorithm. Its input register |r⟩ is prepared in a separate step (which turns out to be surprisingly complex (Plesch et al., 2011; Schuld et al, 2019; Schende et al., 2005). Once the unitary transformation produced its output |s⟩, a separate measurement step determines its result.

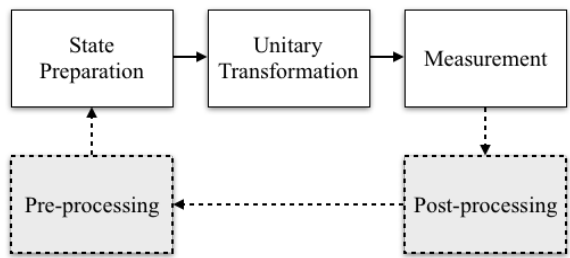

Figure 3: Basis structure of a quantum algorithm.

Optionally, some pre-processing or some post-processing is performed in a classical environment turning the overall algorithm into a hybrid one. Especially, many successful algorithms in a NISQ environment make use of classical processing to reduce the execution time on a quantum computer: the goal is to avoid decoherence and gate faults by spending only a short amount of time on a noisy quantum machine.

One example is a hybrid algorithm called Variational Quantum Eigensolver for determining eigenvalues (Peruzzo et al., 2014). This can be done by using a parameterized quantum algorithm computing and measuring expectation values, which are post-processed on a classical computer. The post-processing consists of a classical optimization step to compute new parameters to minimize the measured expectation values. The significance of this algorithm lies in the meaning of eigenvalues for solving many practical problems (see section 5.2.2).

Another example is the Quantum Approximate Optimization Algorithm (Fhari et al., 2014) that is used to solve combinatorial optimization problems. It computes a state on a quantum machine the expectation values of which relate to values of the cost function to be maximized. The state is computed based on a parameterized quantum algorithm, and these parameters are optimized by classical algorithms in a post-processing step as before. Since many machine learning algorithms require solving optimization problems, the importance of this algorithm is obvious too (see section 5.2.4).

An overview on several fundamental (non-hybrid) algorithms can be found in (Montanro, 2016).

## 2.4 Quantum Software Stack

Programming a quantum computer is supported by a software stack the typical architecture of which is shown in Figure 4. (LaRose, 2019) describes incarnations of this stack by major vendors. Also, section 3 discusses details of some implementations.

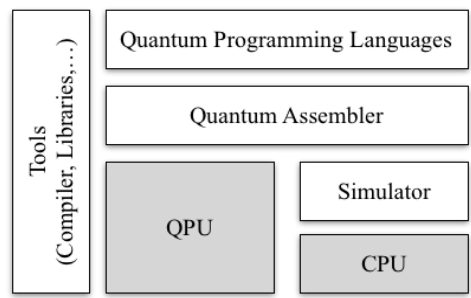

Figure 4: Principle architecture of today's quantum software stack.

The heart of the stack is a quantum assembler: it provides a textual rendering for key unitary transformations that are used to specify a quantum algorithm.

Since a quantum assembler is very low level, quantum programming languages are offered that host the elements of the quantum assembler in a format more familiar to traditional programmers – but still, the assembler flavor is predominant. In addition, functions to connect to quantum machines (a.k.a. quantum processing unit QPU) and simulators etc. are provided.

Quantum programming languages also come with libraries that provide implementations of often used quantum algorithms to be used as subroutines.

A compiler transforms a quantum assembler program into an executable that can be run on a certain QPU. Alternatively, the compiler can transform the quantum assembler into something executable by a simulator on a classical CPU.

## 2.5 Sample Research Questions

The most fundamental question is about a proper engineering discipline for building (hybrid) quantum applications. For example: What development approach should be taken? How do quantum experts interact with software engineers? How are quantum applications tested, debugged?

# 3 QUANTUM AS A SERVICE

Since quantum algorithms promise to speed up known solutions of several hard problems in computer science, research in the field of software development for quantum computing has increased in recent years. In order to achieve speedup against classical algorithms, quantum algorithms exploit certain quantum-specific features such as superposition or entanglement (Jozsa and Linden, 2003). The implementation of quantum algorithms is supported by the quantum software stack as shown in Figure 4. In this section, we give an overview of current tools for the development of quantum software. We further discuss deployment, different service models, and identify open research areas.

## 3.1 Tooling

Several platforms implementing the introduced quantum computing stack have been released in recent years (LaRose, 2019). This includes platforms from quantum computer vendors, such as Qiskit (Qiskit, 2020) from IBM or Forest (PyQuil, 2020) from Rigetti, as well as platforms from third-party vendors such as ProjectQ (Steiger et al., 2018) or XACC (McCaskey et al., 2019).

The quantum algorithms are described by so-called quantum circuits which are structured collections of quantum gates. These gates are unitary transformations on the quantum register (see section 2.3). Each platform provides a universal set of gates that can be used to implement any quantum algorithm. Figure 5 shows a simple example of such a circuit. It uses two qubits (each represented as a horizontal line), both of which are initialized as $|0\rangle$. A classical two-bit register c is used for the results of measurement and depicted as one single line. The Hadamard gate (H), which creates an equal superposition of the two basis states $|0\rangle$ and $|1\rangle$, is applied to the qubit at quantum register position 0. Then, the Controlled Not gate (CNOT) is applied to the qubits at quantum register positions 0 and 1, whereby the former acts as control-bit and a NOT operation is applied to the second qubit iff the control

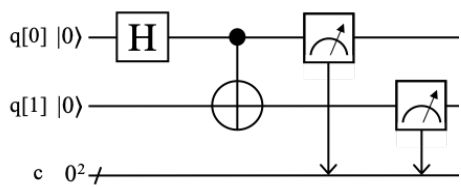

Figure 5 Example of a quantum circuit.

```
1  from SDK import lib
2  # create circuit and add gates
3  circuit = lib.Circuit()
4  circuit.H(0)
5  circuit.CNOT(0, 1)
6  ...
7  # many more
8  ...
9  circuit.measure()
10 # choose QPU
11 backend = lib.getBackend('...')
12 # compile circuit and send to QPU
13 result = lib.execute(circuit,
   backend, shots)
```
Listing 1: Sample code snippet for the creation and execution of a quantum circuit.

qubit is $|1\rangle$. Finally, measurement gates are added to both qubits stating that these qubits will be measured and the resulting values will be stored in the classical bit register.

The different platforms support different quantum programming languages which are embedded in classical host languages, such as PyQuil from Forest embedded in Python, or Qiskit embedded in Python, JavaScript, and Swift. The platforms provide libraries with methods for implementing a quantum circuit. Listing 1 shows a code snippet example of the creation and execution of the circuit from Figure 5. The first line imports the library. Then, a circuit object is created to accumulate the gates in sequential order. Gate H is added to the circuit in line 4 and the CNOT gate is added to the circuit in line 5. Finally, measurement is added to the circuit in line 9. After the circuit is built, a concrete backend is chosen in line 11, which can be either a local simulator, a simulator in the cloud, or a QPU. The execution of the circuit is initiated in line 13. This execute method requires the circuit, the chosen backend, and the number of shots as input. As stated in section 2.2, a quantum algorithm is normally executed multiple times and the number of executions can be configured using the shots parameter.

The circuit is then converted to quantum assembler language by the complier of the respective platform, e.g., to OpenQASM (Cross et al., 2017) for QPUs of IBM, or Quil (Smith et al., 2016) for QPUs of Rigetti. In section 4.4 quantum compilers are introduced in more detail. The compiled code is sent to the selected backend. The execution itself normally is job-based, meaning that it will be stored in a queue before it gets eventually executed. The result, as mentioned before, is a probability distribution of all measured register states and must be interpreted afterwards.

Although the vendor-specific libraries are embedded in high-level programming languages, the implementation of quantum algorithms using the universal sets of gates requires in-depth quantum computing knowledge. Therefore, libraries sometimes already provide subroutines for common quantum algorithms, such as the Variational Quantum Eigensolver, or Quantum Approximate Optimization Algorithm. (LaRose, 2019) compares different libraries with regards to their provided subroutines. However, these subroutines can often not be called without making assumptions about their concrete implementation and the used QPU.

Currently, most platforms are provided by the quantum computer vendors and are, thus, vendor-specific. However, there are also vendor-agnostic approaches, such as ProjectQ or XACC that both are extensible software platforms allowing to write vendor-agnostic source code and run it on different QPUs. Section 4 gives more details on the hardware-agnostic processing of quantum algorithms.

## 3.2 Deployment and Quantum Application as a Service

Several quantum computer vendors provide access to their quantum computers via the cloud. This cloud service model can be called Quantum Computing as a Service (QCaaS) (Rahaman et al., 2015). Also cloud providers, such as Amazon or 1Qbit, have taken QCaaS offerings to their portfolio. The combination of quantum and traditional computing infrastructure is essential for the realization of quantum applications. As already shown in Figure 3, a quantum computer is typically not used on its own but in combination with classical computers: the latter are still needed to store data, pre- and post-process data, handle user interaction, etc. Therefore, the resulting architecture of a quantum application is hybrid consisting of both quantum and classical parts.

The deployment logic of the quantum part is currently included in the source code as shown in Listing 1. For running a quantum application (i) the respective platform has to be installed on a classical computer, (ii) the circuit must be implemented, (iii) the backend has to be selected, and (iv) the circuit must be executed. Therefore, we propose another service model that we call Quantum Application as a Service (QaaS), which is depicted in Figure 6. The QaaS offering wraps all application and deployment logic of a quantum application, including the quantum circuit as well as data pre- and post-processing, and provides an APIs that can then be used for integration with traditional application, e.g., web applications or workflows.

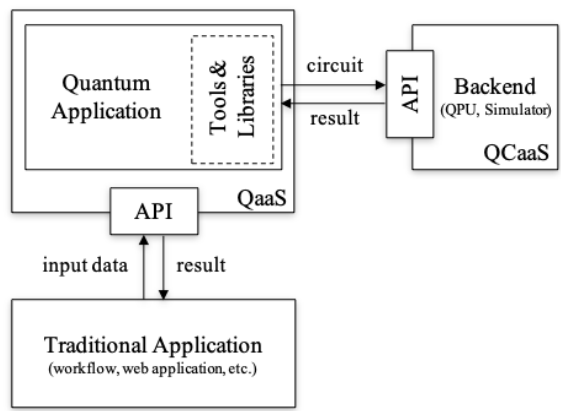

Figure 6: Quantum Algorithm as a Service (QaaS) and Quantum Computing as a Service (QCaaS).

The traditional application passes input data to the API. However, this input data must be properly encoded in order to initialize the quantum register for the following computation (Leymann, 2019). This data encoding, the construction of an appropriate quantum circuit, its compilation, and the deployment is all handled by the service. For the execution of the circuit itself a QCaaS offering can be used. A hardware-agnostic processing of quantum algorithms would also enable the dynamical selection of different QCaaS as further discussed in section 4. The result of this execution is interpreted by the quantum application and finally returned to the traditional application.

This concept would enable to separate quantum applications from traditional applications, particularly with regard to their deployment. Furthermore, the integration of quantum computing features can be eased since QaaS enables to use common technologies of service-based architectures.

## 3.3 Sample Research Questions

To realize the proposed concept, the driving question is: How are hybrid quantum-classical applications deployed? In addition, the integration of quantum applications with traditional applications must be considered. This raises further questions. For example: What are the details of quantum algorithms, and especially their input and output formats? What are efficient encodings of input data? And for which parts of an application can a speedup be achieved?

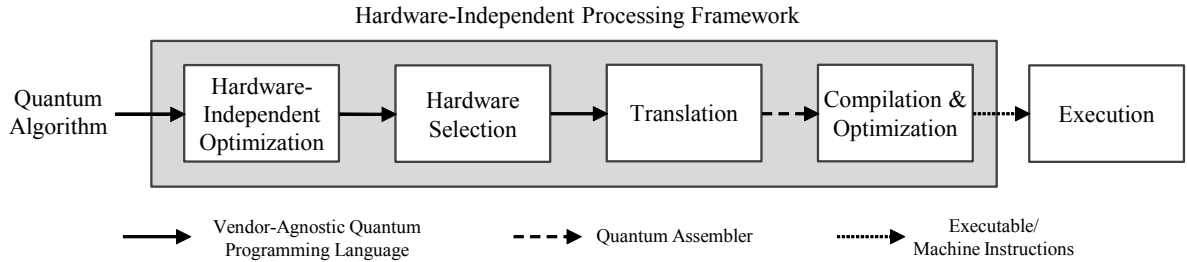

Figure 7: Processing of hardware-independent quantum algorithms

## 4 REMOVING HARDWARE DEPENDENCIES

In this section, we motivate the need for removing the dependencies of quantum algorithms from quantum hardware and vendor-specific quantum programming languages. Afterwards, we present a method for the processing of hardware-independent quantum algorithms. Further, we sketch existing approaches to compile quantum algorithms to executables, optimize them, and show open research questions for selecting and distributing the quantum algorithms over suitable quantum and classical hardware.

### 4.1 Problem

Due to the rapid development and improvement of quantum computers (National Academies, 2019), it is important to keep implementations of quantum algorithms as hardware-independent and portable as possible, to enable the easy exchange of utilized quantum machines. Novel quantum algorithms are mostly specified and published in the abstract quantum circuit representation (Svore et al., 2006). Therefore, to execute them, they must be implemented using the quantum programming language of a specific vendor (see section 3.1). However, the quantum programming languages are not standardized and are usually only supported by a small subset or even only one quantum hardware vendor (LaRose, 2019). Therefore, the implementation of a quantum algorithm utilizing a specific quantum programming language can lead to a vendor lock-in. To circumvent this problem, a standardized, machine-readable, and vendor-agnostic representation for quantum circuits is required, which can be automatically translated into the representations of the different vendor-specific quantum programming languages (see section 2.4).

Furthermore, after specifying a quantum algorithm using a certain quantum programming language, the utilized qubits and gates must be automatically mapped to qubits, gates, and measurements that are provided by the quantum machine to keep them independent of different quantum machines of a specific vendor (Booth Jr, 2012).

### 4.2 Hardware-Independent Processing

In this section, we present a method for the processing of hardware-independent quantum algorithms, which is based on the works of (Häner et al., 2018) and (McCaskey et al., 2020). First, the required steps are presented and afterwards the following sections introduce available research works that can be integrated into the approach and provide an overview of open research questions for the different steps.

The required processing steps for hardware-independent quantum algorithms are sketched in Figure 7. The inputs and outputs of the different steps are depicted by the arrows connecting them. First, the quantum algorithm is defined utilizing a vendor-agnostic quantum programming language, which should be standardized and comprise all relevant parts of quantum algorithms (McCaskey et al., 2020). Then, a hardware-independent optimization can be performed (see section 4.5), which, e.g., deletes unnecessary qubits or gates (Häner et al., 2018).

Based on the optimized quantum algorithm, suitable quantum hardware is selected in the next step. For this, important properties characterizing the quantum algorithm, such as the required number of qubits or the utilized gate set, are retrieved (Suchara et al., 2013). Due to the limited quantum hardware in the NISQ era (Preskill, 2019), this information is important and can be used to select a quantum computer that can successfully execute the quantum algorithm. Furthermore, this selection can be based on different metrics, such as the error-probability, the occurring costs, or the set of vendors that are trusted by the user (McCaskey et al., 2020).

After the selection of the quantum hardware to execute an algorithm, the algorithm must be translated from the vendor-agnostic quantum programming language to the quantum assembler of a vendor that supports the execution on the selected quantum hardware (McCaskey et al., 2020). Next, it can be compiled to an executable for the selected quantum hardware. For this, the available vendors usually provide suitable compilers (see section 4.4) (LaRose, 2019). During the compilation process, hardware-dependent optimizations are performed. Finally, the executable can be deployed and executed on the selected quantum machine (see section 3.2).

### 4.3 NISQ Analyzer

The NISQ Analyzer is a component which analyzes quantum algorithms and extracts the important details, such as the number of required qubits or the utilized gate set (Suchara et al., 2013). Therefore, the quantum algorithm specified in the hardware-independent quantum programming language can be used as an input for the NISQ Analyzer. However, the analysis of quantum algorithms and the precise estimation of resource requirements are difficult problems (Scherer et al., 2017). For example, the required gates for the initial data encoding (Leymann, 2019) or the overhead due to required error correction codes (Laflamme et al., 1996) must be considered. Additionally, the resource requirements for oracle implementations are often ignored but lead to a large overhead that should be noted (Scherer et al., 2017). Thus, tooling support is required that extracts all relevant characteristics of quantum algorithms and provides them to the other components, such as the quantum compiler.

### 4.4 Quantum Compiler

The quantum compiler is in charge of performing the mapping from the quantum assembler representing a quantum algorithm to an executable for a concrete quantum computer (Booth Jr, 2012; Heyfron and Campbell, 2018). The mapping of gates and measurements that are physically implemented by a quantum computer can be performed directly. However, gates and measurements that are not physically available have to be mapped to a "subroutine" consisting of physical gates and measurements (Heyfron and Campbell, 2018). For example, if a measurement using a certain basis is not implemented, the quantum state must be transferred into a basis for which a measurement is provided by the quantum hardware and the measurement must be done in this basis. The utilized subroutines strongly influence the execution time and error probability of the calculation, as they add additional gates and measurements (Steiger et al., 2018). Hence, suited metrics and algorithms to select the required subroutines are important to reduce the overhead of the mapping (see section 4.5). Additionally, the qubits must be mapped to available physical qubits, which influences the quantum algorithm execution as well, due to different characteristics of the qubits, such as decoherence time or connectivity (Zhang et al., 2019). However, the available quantum compilers are mostly vendor-specific (LaRose, 2019), and therefore, compile the quantum algorithm implementations defined in the quantum assembler of a certain vendor to the executable for concrete quantum hardware that is provided by this vendor. Other quantum compilers define their own quantum assembler language to specify quantum algorithms and map them to executables for a certain quantum computer as well (Javadi-Abhari et al., 2015). Thus, the dependency on the vendor- or compiler-specific quantum assembler language cannot be removed by these kinds of quantum compilers. Hence, quantum compilers must be integrated into the approach for processing hardware-independent quantum algorithms (see Figure 7).

### 4.5 Optimization of Quantum Algorithms

Quantum algorithms can be optimized in two ways: (i) hardware-independent or (ii) hardware-dependent (Häner et al., 2018). For the hardware-independent optimization, general optimizations at the quantum circuit level are performed, according to a cost function, such as the circuit size or the circuit depth (Svore et al., 2006). In contrast, hardware-dependent optimization takes hardware-specific characteristics, such as the available gate set of the target quantum computer or the decoherence time of different qubits, into account (Itoko et al., 2020). Hence, this optimization is often combined with the compilation to an executable for a certain quantum computer.

In the following, we sketch some existing works regarding the optimization of quantum algorithms. (Heyfron and Campbell, 2018) propose a quantum compiler that reduces the number of T gates, while using the Clifford + T gate set. They show that the cost of the T gate is much higher than for the other Clifford gates, and therefore, they improve the circuit costs by decreasing the T count. (Itoko et al., 2020) present an approach to improve the hardware-dependent mapping from the utilized qubits and gates in the quantum algorithm to the provided qubits and gates of the quantum computer during the compilation process. (Maslov et al., 2008) propose an approach that is based on templates to reduce the circuit depth, which means the number of gates that are executed in sequence on the qubits. A template is a subroutine that can be used to replace functionally equivalent circuit parts by more efficient ones in terms of different metrics like cost or error probability. Hence, they introduce a method to detect and replace suitable circuit parts with templates.

### 4.6 Sample Research Questions

For the definition and processing of hardware-independent quantum algorithms and the selection of suitable quantum hardware, different research questions must be solved, some of which are presented in the following.

The definition of an abstract hardware-independent quantum programming language is important to remove the hardware dependencies of quantum algorithms. Therefore, sample research questions are: What elements are required to define quantum algorithms? How should suited modeling tooling support look like? What subroutines are important and should be provided as libraries?

To automatically select the best available quantum hardware for a quantum algorithm, suited tooling support must be developed. Hence, open research questions are: What characteristics of quantum algorithms are important for the hardware selection? How can these characteristics be retrieved automatically? What are suited metrics and algorithms for the hardware selection? What are the interesting optimization goals?

The hardware-dependent and -independent optimization of quantum algorithms are especially important in the NISQ era. Therefore, interesting research questions are: What are new or improved optimization algorithms? What data about quantum hardware is relevant for the optimization and how can it be obtained?

By comparing the performance of different quantum compilers, the compiler with the best optimization result or best execution time can be selected. Hence, sample research questions are: What are suited benchmarks for the comparison of quantum compilers? How can the optimality of the compiled executable be verified with respect to different optimization goals, like the number of required gates or the number of fault paths?

## 5 QUANTUM MACHINE LEARING: A USE CASE

Determining how quantum computing can solve problems in machine learning is an active and fast-growing field called quantum machine learning (Schuld, 2015). In this section we give a use case from the digital humanities (Berry, 2012) that shows how quantum machine learning can be applied.

### 5.1 MUSE

The use case presented is from our digital humanities project MUSE (Barzen et al., 2018; MUSE, 2020). It aims at identifying costume patterns in films. Costume patterns are abstract solutions of how to communicate certain stereotypes or character traits by e.g. the use of specific clothes, materials, colors, shapes, or ways of wearing. To determine the conventions that have been developed to communicate for example a sheriff or an outlaw, MUSE developed a method and a corresponding implementation to support the method to capture and analyze costumes occurring in films.

The method consists of five main steps: (1) defining the domain by an ontology, (2) identifying – based on strict criteria – the films having most impact within the domain, (3) capturing all detailed information about costumes in films in the MUSE repository, (4) analyzing this information to determine costumes that achieve a similar effect in communicating with the recipient, and (5) abstracting these similarities to costume patterns (Barzen et al., 2018; Barzen, 2018). This method has been proven to be generic by applying it in our parallel project MUSE4Music (Barzen et al., 2016).

#### 5.1.1 Ontology

To structure costume parameters that have a potential effect on the recipient of a film a detailed ontology was developed (Barzen, 2013). This ontology brings together several taxonomies structuring subparts like types of clothes, materials, function, or condition, as well as relations (e.g. worn above, tucked inside, wrapped around, etc.) on how base elements (e.g. trousers, shirts, boots, etc.) are combined into an overall outfit. The 3151 nodes of the ontology induces the schema of the MUSE repository. The repository facilitates the structured capturing of all relevant information about the films, their characters and their costumes.

### 5.1.2 Data Set

The MUSE data set currently (February 2020) contains more than 4.700 costumes out of 57 films, consisting of more than 26.00 base elements, 57,000 primitives (e.g. collar, sleeves, buttons, etc.), 145.000 colors and 165.000 material selections.

Being part of the open data initiative, this data set is freely available to be used and analyzed (MUSE GitHub, 2020). It provides a well-structured and labelled data set that allows several analysis techniques to be applied. Especially promising are techniques from machine learning like feature extraction, clustering, or classification.

### 5.1.3 Data Analysis

As a first approach to analyze the data to identify those significant elements a costume designer uses to achieve a certain effect, a two-step analysis process was introduced (Falkenthal et al., 2016). The first step applies data mining techniques – mainly association rule mining – to determine hypotheses about which elements are used to communicate a certain stereotype, for example. The second step aims at refining and verifying such hypotheses by using online analytical processing (OLAP) techniques (Falkenthal et al., 2015) to identify indicators for costume patters.

To improve the process of building hypotheses that hint to potential costume patterns we are currently extending the analysis of the MUSE data by various techniques from machine learning. Each costume has several properties that describe it in detail. Simply mapping each property of a costume to a feature, the resulting feature space would be of huge dimension. Therefore, feature extraction, namely principle component analysis (PCA), is applied to reduce the dimension of the feature space without losing important information (see section 5.2.2). To group those costumes together that achieve the same effect different cluster algorithms are applied and evaluated (see section 5.2.4). As there are new costumes stored at the database frequently the usage of classification algorithms is investigated (see section 5.2.5) to enable that these costumes get classified as part of the right pattern identified before.

Currently, this approach is implemented on a classical computer with classical machine learning algorithms. But since quantum computing can contribute to solve several problems in machine learning – as shown in the following section – it is promising to improve the approach by not only using classical computer but to also use the potentials offered by quantum computers (Barzen et al., 2020).

## 5.2 Potential Improvements

Several machine learning algorithms require the computation of eigenvalues or apply kernel functions: these algorithms should benefit from improvements in the quantum domain. Many machine learning algorithms are based on optimization, i.e. improvements in this area like Quantum Approximate Optimization Algorithm QAOA should imply improvements of those machine learning algorithms.

Whether or not such improvements materialize is discussed in several papers that compare sample classical and quantum machine learning algorithms, e.g. (Biamonte et al., 2017; Ciliberto et al., 2018; Havenstein et al., 2018).

### 5.2.1 Data Preparation

The data captured in MUSE are categorical data mostly. Since most machine learning algorithms assume numerical data, such categorical data must be transformed accordingly: this is a complex problem.

For example, the different colors of pieces of clothes could be assigned to integer numbers. But the resulting integers have no metrical meaning as required by several machine learning algorithms. Instead of this, we exploited the taxonomy that structures all of our categorical data by applying the Wu and Palmer metric (Wu et al., 1994) to derive distances between categorial data. In addition, we used word embeddings based on restricted Boltzmann machines (Hinton, 2012).

As described above, costumes have a large number of features, thus, this number must be reduced to become tractable. We experiment with feature extraction based on restricted Boltzmann machines (Hinton et al., 2006) as well as with principal component analysis (see section 5.2.2). Feature selection based on deep Boltzmann machines (Taherkhania et al., 2018) may also be used.

### 5.2.2 Eigenvalues

Principal component analysis strives towards combining several features into a single feature with high variance, thus, reducing the number of features. For example, in Figure 8 the data set shown has high variance in the A axis, but low variance in the B axis, i.e. A is a principal component. Consequently, the X and Y features of the data points are used to compute A values as a new feature, reducing the two features X and Y into a single feature A.

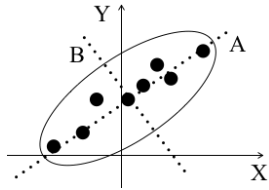

Figure 8: Principal component of a data set.

The heart of this analysis is the calculation of the half axes and their lengths of the ellipse "best" surrounding the data set. This is done by determining the eigenvalues of the matrix representing the ellipse. Computing eigenvalues can be done on a quantum computer much faster than classically by means of quantum phase estimation and variational quantum eigensolvers. Thus, Quantum principal component analysis (Lloyd et al., 2014) is an algorithm we will use in our use case.

### 5.2.3 Quantum Boltzmann Machines

(Zhang et al., 2015) provided a quantum algorithm of a quantum restricted Boltzmann machine. In a use case, it has shown performance superior to a classical restricted Boltzmann machine.

Similarly, (Amin et al., 2018) described an approach for both, quantum Boltzmann machines as well as quantum restricted Boltzmann machines. They report that the quantum restricted Boltzmann machine outperforms the classical restricted Boltzmann machine for small size examples.

Thus, quantum Boltzmann machines are candidates for our use case, especially because they can be exploited in clustering and classification tasks.

### 5.2.4 Clustering

Several quantum clustering algorithms and their improvements over classical algorithms are presented in (Aimeur et al., 2007). Since clustering can be achieved by solving Maximum Cut problems, some attention has been paid to solve MaxCut on quantum computers.

For example, (Crooks, 2018) as well as (Zhou et al., 2019) use QAOA to solve MaxCut problems on NISQ machines. A similar implementation on a Rigetti quantum computer has been described by (Otterbach et al., 2017)

Thus, quantum clustering is promising.

### 5.2.5 Classification

Support vector machines (SVM) are established classifiers. (Rebentrost et al., 2014) introduce quantum support vector machines and show an exponential speedup in many situations.

(Schuld et al., 2014) present a quantum version of the k-nearest neighbour algorithm, and an implementation of a classifier on IBM Quantum Experience (Schuld et al., 2017). A hybrid classifier has been introduced by (Schuld et al., 2018).

The use of kernels in machine learning is well-established (Hofman et al., 2008), and kernels are used in case non-linear separable data must be classified. A hybrid classifier that makes use of kernels is given in (Schuld et al., 2019). (Ghobadi et al., 2019) describe classically intractable kernels for use even on NISQ machines.

Thus, quantum classifiers are promising.

## 5.3 Quantum Humanities

As stressed by the presented use case there are promising application areas for quantum computing not only in industry or natural science but also in the humanities. We coined the term quantum humanities for using quantum computing to solve problems in this domain (Barzen et al., 2019). It aims at exploiting the potentials offered by quantum computers in the digital humanities and raise research questions and describe problems that may benefit from applying quantum computers.

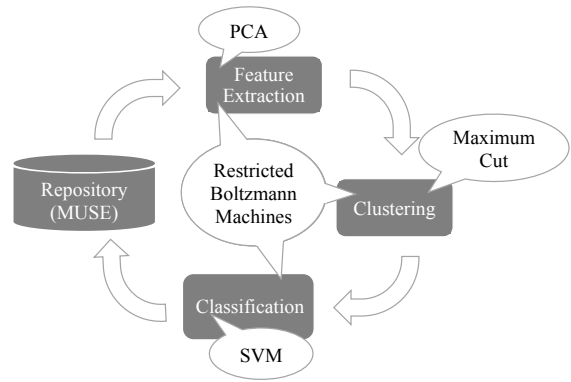

Figure 9: MUSE data analysis.

Figure 9 shows the process and algorithms used to analyze the MUSE data. Its application provides a first feasibility study in the domain of quantum humanities. Furthermore, it derives knowledge for researchers as well as components reusable in other domains. Sharing knowledge with other researchers about solving problems with quantum computers is right at the core of the vision of quantum humanities. Therefore, a pattern language for quantum computing as introduced in (Leymann, 2019) can provide reusable knowledge that enables interested parties that are not too familiar with the algorithmic or mathematical aspects of quantum computing to also participate at the potentials offered by quantum computers. In order to provide not only reusable knowledge, but also an advanced platform that supports several steps in the work with quantum computers (Leymann et al., 2019), section 6 outlines the collaborative quantum platform we are currently building.

## 5.4 Sample Research Questions

The most essential and fundamental question for quantum humanities is to evaluate which existing and new problems from the humanities can be addressed by quantum computers. Especially, which problems are best solved by classical, hybrid, or quantum algorithms? Beside speedup, which algorithms result in higher precision?

Which language allows to communicate between many disciplines (e.g. mathematics, physics, computer science, and the different areas from the humanities)? Are there completely new questions from the humanities that are only addressable based on a quantum computer?

## 6 COLABORATIVE QUANTUM APPLICATION PLATFORM

Driven by the continuous improvement of quantum hardware, specialists in various fields have developed new quantum algorithms and applications in recent years. The use of these quantum applications requires in-depth knowledge of theory and practice, which is often lacking in small and medium-sized companies. A major challenge today is to facilitate the transfer of knowledge between research and practice to identify and fully exploit the potential of new emerging technologies. To prepare a body of knowledge for quantum computing reasonably and make it usable for different stakeholders, a collaborative platform where all participants come together is essential (Leymann et al., 2019). For this purpose, the quantum application platform must cover the entire process from the development of quantum algorithms to their implementation and execution. The diversity of stakeholders and their different objectives lead to a variety of requirements for such a quantum platform.

Building upon the stakeholders identified by (Leymann et al., 2019), we firstly identify key entities, which serve as an anchor for the knowledge on a quantum platform, secondly identify essential requirements for their expedient implementation and, finally, show a general extendable architecture for a collaborative quantum software platform.

## 6.1 Key Entities

To foster a clear structuring of the knowledge created on a quantum software platform the following key entities can be used. They allow different experts to hook into the platform and enables to share and contribute knowledge.

**Quantum Algorithm**: As mentioned before, quantum algorithms are developed and specified typically by experts with in-depth quantum physics background. Thus, for a quantum software platform it is essential to capture quantum algorithms as artifacts. Besides generally sharing them, further valuable information can be attached to quantum algorithms, such as discussion among experts regarding resource consumption of an algorithm, its speedup against classical algorithms, or its applicability to NISQ computers.

**Algorithm Implementation**: Besides the representation of quantum algorithms in their conceptual form, i.e., as mathematical formulas or abstract circuits, the heterogeneous field of quantum hardware demands to capture vendor- and even hardware-specific implementations of quantum algorithms. This is because, implementations for a particular quantum computer offering of a vendor requires the use of a vendor-specific SDK. Thus, implementations of an algorithm for quantum computers offered by different vendors ends up in different code or even the usage of completely different quantum programming languages. Thus, enabling sharing of different algorithm implementations on a quantum software platform stimulates knowledge transfer and reduces ramp-up especially for unexperienced users.

**Data Transformator**: Since quantum algorithms rely on the manipulation of quantum states they do not operate directly on data as represented in classical software. Instead, the data to be processed must be encoded in such a way that they can be prepared into a quantum register. Different problem classes such as clustering or classification of data have specific requirements for the data to be processed. It can be of great benefit to identify general transformation and coding strategies for relevant problem classes. Such strategies can then be represented and discussed on the platform as data transformators.

**Hybrid Quantum Application**: Since only the quantum parts of an algorithm are executed on a quantum computer, they must be delivered together with classical software parts that run on classical computers. To exploit the full potential of quantum algorithms, they often have to be properly integrated into an already running system landscape, which includes proper data preparation and transformation. This is why solutions that are rolled out in practice are typically hybrid quantum applications (see section 3.2). Therefore, knowledge transfer about applicable software solutions for particular use cases at hand is bound to hybrid quantum applications.

**Quantum Pattern**: Software patterns are widely used to capture proven solution principles for recurring problems in many fields in computer science. Thus, quantum patterns seem to be a promising approach to also capture proven solutions regarding the design of quantum algorithms, their implementation and integration in existing systems. First patterns for developing quantum algorithms have already been published (Leymann, 2019).

## 6.2 Requirements

The essential challenge to create and provide a reasonable body of knowledge on quantum algorithms and applications involves the collaboration among several stakeholders. In contrast to traditional software engineering, quantum algorithms are typically not specified by computer scientist rather than by quantum physicists. Furthermore, to understand and implement those algorithms a different mindset is required because the key buildings blocks of algorithms are no longer loops, conditions, or procedure calls but quantum states and their manipulation via unitary operators.

By involving all participants identified by (Leymann et al., 2019) in the platform, added value can be created, both for experienced quantum specialists and inexperienced customers. For this the following listed requirements must be met.

**Knowledge Access:** Often only certain specialists and scientists have the required expertise for developing quantum algorithms and their implementation. To identify and exploit the use cases of quantum computing in practice, companies must be empowered to gather knowledge and to exchange with experts (developer, service provider, consultants, and so on) (Mohseni et al., 2017). Additionally, due to the high level of research activities in this area, the exchange between experts is important in order to share and discuss new findings with the community at an early stage.

**Best Practices for Quantum Algorithm Development:** The development of new algorithms requires in-depth knowledge and expertise in theory and practice. Documented, reusable best practices for recurring problems, i.e. patterns, can support and guide people in the development of new quantum algorithms.

**Decision-Support for Quantum Applications and Vendors:** A two-stage decision-support is required to identify appropriate solutions for real-world use cases. First, quantum algorithms that prove to provide a solution for a given problem have to be identified. Second, the appropriate implementation and quantum hardware have to be selected for integration and execution. For the second stage the resource consumption of algorithms and implementations on different quantum hardware are of main interest (see section 4.2).

**Vendor-Agnostic Usage of Quantum Hardware:** Currently, various algorithm implementations from different vendors are available via proprietary SDKs that have been developed specifically for their hardware. To avoid vendor lock-in the quantum algorithm must be portable between different vendors which can be achieved by a standardized quantum programming language (see section 3.1 and 4.2).

**Data Transformation for Quantum Algorithms:** Especially for machine learning and artificial intelligence data of sufficient quality is essential. This applies to both, classical and quantum algorithms. Such data have to be made available and respectively encoded for the quantum algorithm (Mitarai et al., 2019).

**Quantum Application as a Service (QaaS):** The hybrid architecture of quantum applications consisting of classical and quantum parts increases the complexity of their deployment. Quantum applications provided "as a Service" via a self-service portal ease the utilization of the new technology (see section 3.2).

## 6.3 Architecture

In Figure 10 the architecture of the collaborative quantum software platform is depicted. In essence, the platform consists of two parts: The analysis and development platform as depicted on the left of the figure for collecting, discussing, analyzing, and sharing knowledge, and the marketplace as depicted on the right that offers solutions in the form of quantum applications and consulting services.

The analysis and development platform addresses the needs of specialists and researchers in the field of quantum computing and software engineering. In a first step, knowledge in the form of publications, software artifacts, datasets, or web content can be placed on the platform – either manually via a user interface or automatically using a crawler. This knowledge can originate from various sources, such as arXiv.org or github.com. In a first step it can be stored as raw data in the QAlgo & data content store. Content of interest has to be extracted from these raw data, such as a quantum algorithm described in a journal article. To facilitate collaboration among different disciplines and to create a common understanding, the representation of quantum circuits and mathematical expressions must be normalized. A qualified description of the knowledge artifact with metadata is also essential to find and link relevant knowledge. Therefore, metadata formats must be normalized and enriched. The knowledge artifacts are then stored and provided via an expert portal to specialists and scientists and via a customer portal to users looking for solutions for their use cases and the community of interested people.

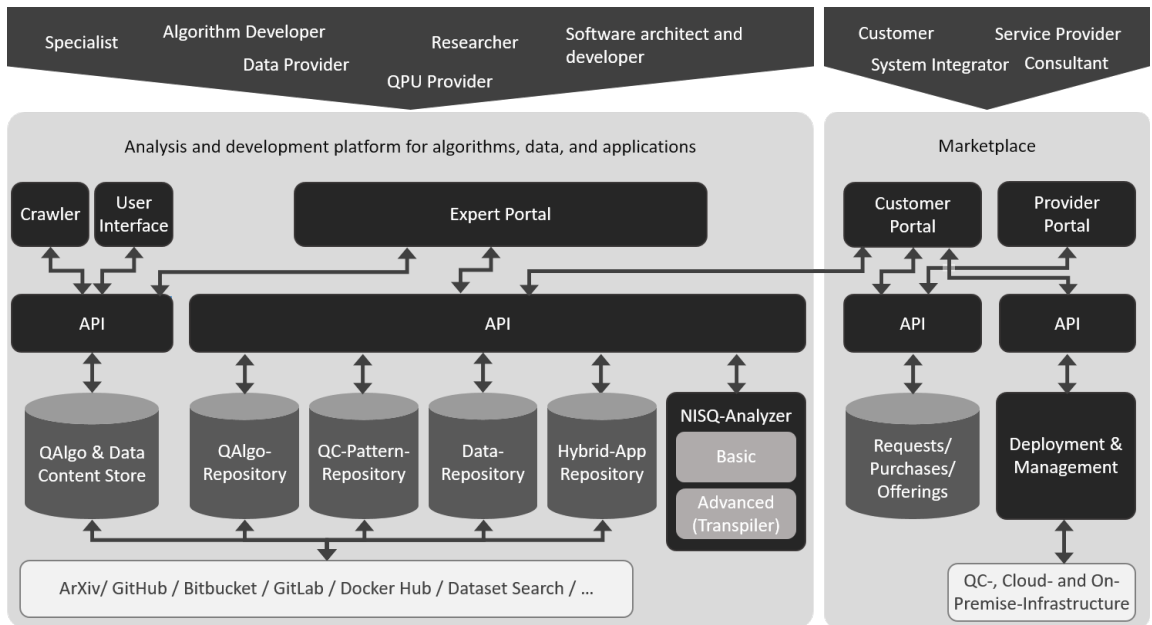

Figure 10: Architecture for a collaborative quantum software platform.

Specialists and scientists can discuss, evaluate, and improve the different key entities on the platform. Algorithms and their implementations can be linked and evaluated based on defined metrics using the NISQ-Analyzer (see section 4.3). Identified best practices, e.g., for creating entanglement, can then be stored as quantum patterns in a Quantum Computing Pattern Repository. These patterns ease the development of new algorithms as they provide proven solutions for frequently occurring problems at the design of quantum algorithms. Patterns solving specific problems can then be combine and applied for realizing a broader use case (Falkenthal et al., 2014; Falkenthal et al., 2017). However, best practices are not only relevant for the development, but also for data preparation as input for quantum algorithms and the integration of quantum algorithms with classical applications. Data preparation is essential, and must especially be considered in the NISQ era.

Since most quantum algorithms are hybrid algorithms, execution of quantum applications means a distributed deployment of hybrid quantum applications among classical and quantum hardware. Such applications can be stored for reuse in the Hybrid-App-Repository. For the quantum part, the quantum computer vendor and more specific a single QPU has to be selected, depending on the QPU properties, the algorithm implementation, and the input data. The platform automates this selection and provides a vendor-agnostic access to quantum hardware. For the deployment, technologies for classical computing are evaluated to provide an integrated deployment automation toolchain. Standards such as the Topology and Orchestration Specification for Cloud Applications (TOSCA) (OASIS, 2019) have been developed precisely for this purpose to enable portability, interoperability, and the distribution across different environments (Saatkamp et al., 2017; Saatkamp et al., 2019). Thus, TOSCA as an international standard offers good foundation for an integration of classical and quantum deployment.

While the expert portal is tailored to provide a sufficient user interface and toolchain addressing the needs of quantum computing experts the marketplace on the right of Figure 10 enables service providers and further stakeholders, such as consultants, to offer solutions. Customers can place requests for solutions for certain problems or use cases at hand. It is further intended to also allow consulting services to be offered in addition to hybrid quantum applications and their deployments. This means that also business models besides the development and distribution is enabled by the interplay of the marketplace and the analysis and development platform. For example, hybrid quantum applications can be provided as a Service, which is enabled through the automated deployment capabilities by means of a TOSCA orchestrator such as OpenTOSCA (Binz et al, 2013; OpenTOSCA, 2020) or Cloudify (Cloudify, 2020). Further, the selection of quantum algorithms fitting to specific constraints of quantum hardware can be supported by the NISQ-Analyzer and the discussions of experts. With the help of the marketplace, knowledge and software artifacts such as quantum algorithm implementations and hybrid quantum applications can be monetized. Every turnover on the platform leads to incentives for participating experts to make further knowledge available on the platform.

## 6.4 Sample Research Questions

The platform provides the basis for the technical realization of the research questions already discussed. However, further questions are raised: What are best practices for data preparation as input for quantum algorithms? What are best practices for integrating quantum algorithms with classical applications? How to combine the best practices in quantum computing with other domains such as cloud computing? Which metadata is required to adequately describe the key entities on the platform?

## 7 CONCLUSIONS

New possibilities to solve classically intractable problems based on quantum computing is at the horizon. Quantum computers appear as part of the cloud infrastructure, and based on the hybrid nature of quantum-based applications, cloud computing techniques will contribute to the discipline of building them. Lots of new research questions appeared.

We are about to build the collaborative quantum application platform, and exploit it for several use cases, especially in the area of machine learning. A pattern language for quantum computing is under construction. Research on the removal of hardware dependencies including deployment of hybrid quantum applications is ongoing.

## ACKNOWLEDGEMENTS

We are grateful to Marie Salm and Manuela Weigold for discussing several subjects of this paper. Also, our thanks go to Daniel Fink, Marcel Messer and Philipp Wundrack for their valuable input and implementing several aspects sketched here.

This work was partially funded by the BMWi project PlanQK (01MK20005N) and the university of Stuttgart funded Terra Incognita project Quantum Humanities.